\documentclass[12pt,a4paper,fleqn]{article}
\usepackage{amsmath}

\begin{document}

\title{\textbf{On integrability of one third-order nonlinear evolution equation}}
\author{\textsc{S.~Yu.~Sakovich\bigskip}\\{\footnotesize Institute of Physics,
National Academy of Sciences,\vspace {-6pt}}\\{\footnotesize
220072 Minsk, Belarus; sakovich@dragon.bas-net.by}}
\date{}
\maketitle

\begin{abstract}
We study one third-order nonlinear evolution equation, recently introduced by
Chou and Qu in a problem of plane curve motions, and find its transformation
to the modified Korteweg--de~Vries equation, its zero-curvature representation
with an essential parameter, and its second-order recursion operator.
\end{abstract}

\section{Introduction}

In their recent work on the motions of plane curves \cite{CQ}, Chou and Qu
found the following new third-order nonlinear evolution equation:%
\begin{equation}
u_{t}=\tfrac{1}{2}\left(  \left(  u_{xx}+u\right)  ^{-2}\right)  _{x}.
\label{cqe}%
\end{equation}
\textquotedblleft We do not know if this equation arises from the AKNS- or the
WKI-scheme\textquotedblright, wrote Chou and Qu in \cite{CQ}.

In the present paper, we study integrability of (\ref{cqe}). In
Section~\ref{s2}, we find a chain of Miura-type transformations, which relates
the equation (\ref{cqe}) with the modified Korteweg--de~Vries equation (mKdV).
In Section~\ref{s3}, using the obtained transformations and the well-known
zero-curvature representation (ZCR) of the mKdV, we derive a complicated
nontrivial ZCR of (\ref{cqe}), which turns out to be neither AKNS- nor
WKI-type one; and then we prove that simpler ZCRs of the equation (\ref{cqe})
are trivial. In Section~\ref{s4}, we derive a second-order recursion operator
of (\ref{cqe}) from the obtained ZCR. Section~\ref{s5} gives some concluding remarks.

\section{Transformation to the mKdV\label{s2}}

Let us try to transform the equation (\ref{cqe}) into one of the well-known
integrable equations. We do it, following the way described in \cite{Sak1};
further details on Miura-type transformations of scalar evolution equations
can be found in \cite{Sak2}.

First, we try to relate the equation (\ref{cqe}) with an evolution equation of
the form%
\begin{equation}
v_{t}=v^{3}v_{xxx}+g\left(  v,v_{x},v_{xx}\right)  \label{v3g}%
\end{equation}
by a transformation of the type%
\begin{equation}
v\left(  x,t\right)  =a\left(  u,u_{x},\ldots,u_{x\ldots x}\right)  .
\label{gmt}%
\end{equation}
If there exists a transformation (\ref{gmt}) between an evolution equation%
\begin{equation}
u_{t}=f\left(  u,u_{x},u_{xx},u_{xxx}\right)  :\quad\partial f/\partial
u_{xxx}\neq\mathrm{constant} \label{ncs}%
\end{equation}
and an equation of the form (\ref{v3g}), then necessarily%
\begin{equation}
a=\left(  \partial f/\partial u_{xxx}\right)  ^{1/3}. \label{nfa}%
\end{equation}
Applying the transformation%
\begin{equation}
\left(  x,t,u\left(  x,t\right)  \right)  \mapsto\left(  x,t,v\left(
x,t\right)  \right)  :\quad v=-\left(  u_{xx}+u\right)  ^{-1} \label{uvt}%
\end{equation}
to the equation (\ref{cqe}), we find that (\ref{uvt}) really works and relates
(\ref{cqe}) with the equation%
\begin{equation}
v_{t}=v^{3}v_{xxx}+3v^{2}v_{x}v_{xx}+v^{3}v_{x}. \label{v3e}%
\end{equation}

Second, we notice that (\ref{v3e}) can be written in the form%
\begin{equation}
v_{t}=v^{2}\left(  vv_{xx}+v_{x}^{2}+\tfrac{1}{2}v^{2}\right)  _{x}.
\label{v2x}%
\end{equation}
Owing to this property, the equation (\ref{v3e}) admits the transformation%
\begin{equation}
\left(  y,t,w\left(  y,t\right)  \right)  \mapsto\left(  x,t,v\left(
x,t\right)  \right)  :\quad x=w,\quad v=w_{y}, \label{ibr}%
\end{equation}
which turns out to relate (\ref{v3e}) with%
\begin{equation}
w_{t}=w_{yyy}+\tfrac{1}{2}w_{y}^{3}. \label{pmk}%
\end{equation}

And, third, we make the transformation%
\begin{equation}
\left(  y,t,w\left(  y,t\right)  \right)  \mapsto\left(  y,t,z\left(
y,t\right)  \right)  :\quad z=w_{y} \label{pot}%
\end{equation}
of (\ref{pmk}) to the mKdV%
\begin{equation}
z_{t}=z_{yyy}+\tfrac{3}{2}z^{2}z_{y}, \label{mkv}%
\end{equation}
for convenience in what follows, because $z\left(  y,t\right)  =v\left(
x,t\right)  $.

\section{Zero-curvature representation\label{s3}}

\subsection{Transformation of the mKdV's ZCR}

Using the chain of transformations (\ref{uvt}), (\ref{ibr}) and (\ref{pot}),
we can derive a Lax pair for the equation (\ref{cqe}), in the form of a ZCR
containing an essential parameter, from the following well-known ZCR of the
mKdV (\ref{mkv}) \cite{Wad1,Wad2}:%
\begin{equation}
\Phi_{y}=A\Phi,\quad\Phi_{t}=B\Phi,\quad D_{t}A=D_{y}B-\left[  A,B\right]
\label{wsp}%
\end{equation}
with%
\begin{align}
A  &  =%
\begin{pmatrix}
\alpha & \frac{\mathrm{i}}{2}z\smallskip\\
\frac{\mathrm{i}}{2}z & -\alpha
\end{pmatrix}
,\label{wma}\\
B  &  =%
\begin{pmatrix}
\frac{1}{2}\alpha z^{2}+4\alpha^{3} & \frac{\mathrm{i}}{2}z_{yy}%
+\frac{\mathrm{i}}{4}z^{3}+\mathrm{i}\alpha z_{y}+2\mathrm{i}\alpha
^{2}z\smallskip\\
\frac{\mathrm{i}}{2}z_{yy}+\frac{\mathrm{i}}{4}z^{3}-\mathrm{i}\alpha
z_{y}+2\mathrm{i}\alpha^{2}z & -\frac{1}{2}\alpha z^{2}-4\alpha^{3}%
\end{pmatrix}
, \label{wmb}%
\end{align}
where $\Phi\left(  y,t\right)  $ is a two-component column, $D_{t}$ and
$D_{y}$ stand for the total derivatives, $\left[  A,B\right]  $ denotes the
matrix commutator, and $\alpha$ is a parameter.

First of all, we obtain a ZCR for the equation (\ref{v3e}) through the
transformations (\ref{ibr}) and (\ref{pot}). Introducing the column $\Psi
:\Psi\left(  x,t\right)  =\Phi\left(  y,t\right)  $, we have $\Phi_{y}%
=z\Psi_{x}$, which allows to rewrite the equation $\Phi_{y}=A\Phi$ as%
\begin{equation}
\Psi_{x}=X\Psi, \label{psx}%
\end{equation}
where $X=z^{-1}A$ after substitution of $z\left(  y,t\right)  =v\left(
x,t\right)  $,%
\begin{equation}
X=%
\begin{pmatrix}
\alpha v^{-1} & \frac{\mathrm{i}}{2}\smallskip\\
\frac{\mathrm{i}}{2} & -\alpha v^{-1}%
\end{pmatrix}
. \label{mxv}%
\end{equation}
The equation $\Phi_{t}=B\Phi$, due to $\Phi_{t}=\Psi_{t}+w_{t}\Psi_{x}$ and
$w_{t}=z_{yy}+\frac{1}{2}z^{3}$, leads to%
\begin{equation}
\Psi_{t}=T\Psi, \label{pst}%
\end{equation}
where $T=B-\left(  z^{-1}z_{yy}+\frac{1}{2}z^{2}\right)  A$ after substitution
of $z=v$, $z_{y}=vv_{x}$ and $z_{yy}=v^{2}v_{xx}+vv_{x}^{2}$,%
\begin{equation}
T=%
\begin{pmatrix}
-\alpha\left(  vv_{xx}+v_{x}^{2}\right)  +4\alpha^{3} & \mathrm{i}\alpha
vv_{x}+2\mathrm{i}\alpha^{2}v\smallskip\\
-\mathrm{i}\alpha vv_{x}+2\mathrm{i}\alpha^{2}v & \alpha\left(  vv_{xx}%
+v_{x}^{2}\right)  -4\alpha^{3}%
\end{pmatrix}
. \label{mtv}%
\end{equation}
It is easy to check that the compatibility condition%
\begin{equation}
D_{t}X=D_{x}T-\left[  X,T\right]  \label{cxt}%
\end{equation}
of the equations (\ref{psx}) and (\ref{pst}), with the matrices $X$
(\ref{mxv}) and $T$ (\ref{mtv}), determines exactly the equation (\ref{v3e}).

Then we can use the transformation (\ref{uvt}). Substituting $v=-\left(
u_{xx}+u\right)  ^{-1}$ into $X$ (\ref{mxv}) and $T$ (\ref{mtv}), we obtain a
ZCR of the equation (\ref{cqe}), in the sense that (\ref{cxt}) is satisfied by
any solution of (\ref{cqe}). This ZCR, however, determines not the equation
(\ref{cqe}) itself, but a differential prolongation of (\ref{cqe}),%
\begin{equation}
u_{xxt}+u_{t}=\tfrac{1}{2}\left(  \left(  u_{xx}+u\right)  ^{-2}\right)
_{xxx}+\tfrac{1}{2}\left(  \left(  u_{xx}+u\right)  ^{-2}\right)  _{x},
\label{pro}%
\end{equation}
due to the structure of the transformed matrix $X$,%
\begin{equation}
X=%
\begin{pmatrix}
-\alpha\left(  u_{xx}+u\right)  & \frac{\mathrm{i}}{2}\smallskip\\
\frac{\mathrm{i}}{2} & \alpha\left(  u_{xx}+u\right)
\end{pmatrix}
. \label{tmx}%
\end{equation}

The situation can be improved by a linear transformation of the auxiliary
vector function $\Psi$,%
\begin{equation}
\Psi\mapsto G\Psi,\quad\det G\neq0, \label{lin}%
\end{equation}
which generates a gauge transformation of $X$ and $T$,%
\begin{equation}
X\mapsto GXG^{-1}+\left(  D_{x}G\right)  G^{-1},\quad T\mapsto GTG^{-1}%
+\left(  D_{t}G\right)  G^{-1}. \label{gxt}%
\end{equation}
The choice of%
\begin{equation}
G=%
\begin{pmatrix}
\exp\left(  \alpha u_{x}\right)  & 0\smallskip\\
0 & \exp\left(  -\alpha u_{x}\right)
\end{pmatrix}
\label{cmg}%
\end{equation}
leads through (\ref{gxt}) to the following gauge-transformed matrix $X$, which
does not contain $u_{xx}$:%
\begin{equation}
X=%
\begin{pmatrix}
-\alpha u & \frac{\mathrm{i}}{2}\exp\left(  2\alpha u_{x}\right)  \smallskip\\
\frac{\mathrm{i}}{2}\exp\left(  -2\alpha u_{x}\right)  & \alpha u
\end{pmatrix}
. \label{fmx}%
\end{equation}
Note that $u$ and $u_{x}$ are separated in (\ref{fmx}), and a ZCR with such a
matrix $X$ can determine an evolution equation exactly.

Now, from (\ref{mtv}), (\ref{uvt}), (\ref{gxt}) and (\ref{cmg}), we obtain the
following matrix $T$, where $u_{t}$ and $u_{xt}$ have been expressed through
(\ref{cqe}) in terms of $x$-derivatives of $u$:%
\begin{equation}
T=%
\begin{pmatrix}
4\alpha^{3} & T_{12}\smallskip\\
T_{21} & -4\alpha^{3}%
\end{pmatrix}
\label{fmt}%
\end{equation}
with%
\begin{align}
T_{12}  &  =-\mathrm{i}\alpha\exp\left(  2\alpha u_{x}\right)  \left(
\frac{2\alpha}{u_{xx}+u}+\frac{u_{xxx}+u_{x}}{\left(  u_{xx}+u\right)  ^{3}%
}\right)  ,\label{t12}\\
T_{21}  &  =\mathrm{i}\alpha\exp\left(  -2\alpha u_{x}\right)  \left(
-\frac{2\alpha}{u_{xx}+u}+\frac{u_{xxx}+u_{x}}{\left(  u_{xx}+u\right)  ^{3}%
}\right)  . \label{t21}%
\end{align}
It is easy to check that the matrices $X$ (\ref{fmx}) and $T$ (\ref{fmt}%
)--(\ref{t21}) constitute a ZCR of (\ref{cqe}), in the sense that the
condition (\ref{cxt}) with these matrices determines exactly the equation
(\ref{cqe}).

\subsection{Simpler ZCRs are trivial}

The obtained ZCR of (\ref{cqe}) is characterized by the complicated matrix $X$
(\ref{fmx}) containing $u_{x}$. Does the equation (\ref{cqe}) admit any
simpler ZCR, with $X=X\left(  x,t,u\right)  $, of any dimension $n\times n$?
This problem can be solved by direct analysis of the condition (\ref{cxt}).

Substituting $X=X\left(  x,t,u\right)  $ and $T=T\left(  x,t,u,u_{x}%
,u_{xx}\right)  $ into (\ref{cxt}) and replacing $u_{t}$ by the right-hand
side of (\ref{cqe}), we obtain the following condition, which must be an
identity, not an ordinary differential equation restricting solutions of
(\ref{cqe}):%
\begin{equation}
X_{t}-\frac{u_{xxx}+u_{x}}{\left(  u_{xx}+u\right)  ^{3}}X_{u}=D_{x}T-\left[
X,T\right]  \label{ide}%
\end{equation}
(here and below, subscripts denote derivatives, like $T_{u_{x}}=\partial
T/\partial u_{x}$). Applying $\partial/\partial u_{xxx}$ and $\partial
/\partial u_{xx}$ to the identity (\ref{ide}), we obtain, respectively,%
\begin{align}
T_{u_{xx}}  &  =-\left(  u_{xx}+u\right)  ^{-3}X_{u},\label{tu2}\\
T_{u_{x}}  &  =\left(  u_{xx}+u\right)  ^{-3}\left(  D_{x}X_{u}-\left[
X,X_{u}\right]  \right)  . \label{tu1}%
\end{align}
The compatibility condition $\left(  T_{u_{xx}}\right)  _{u_{x}}=\left(
T_{u_{x}}\right)  _{u_{xx}}$ for (\ref{tu2}) and (\ref{tu1}) is $D_{x}%
X_{u}=\left[  X,X_{u}\right]  $, which is equivalent to%
\begin{equation}
X=P\left(  x,t\right)  u+Q\left(  x,t\right)  :\quad P_{x}=\left[  Q,P\right]
. \label{xpq}%
\end{equation}

Now, we make use of gauge transformations (\ref{gxt}) with $G=G\left(
x,t\right)  $, choose $G$ to be any solution with $\det G\neq0$ of the system
of ordinary differential equations $G_{x}=-GQ$, and thus set $Q=0$ and
$P=P\left(  t\right)  $ in the gauge-transformed matrix $X$ (\ref{xpq}). Then,
$T_{u}=T_{u_{xx}}$ follows from $\partial/\partial u_{x}$ of (\ref{ide}), and
this leads through the identity (\ref{ide}) to%
\begin{equation}
T=\tfrac{1}{2}P\left(  t\right)  \left(  u_{xx}+u\right)  ^{-2}+K\left(
t\right)  :\quad P_{t}=\left[  K,P\right]  . \label{tpk}%
\end{equation}

Finally, we make $K=0$ by a gauge transformation (\ref{gxt}) with $G=G\left(
t\right)  $ satisfying $G_{t}=-GK$ and $\det G\neq0$, and thus obtain%
\begin{equation}
X=Pu,\quad T=\tfrac{1}{2}P\left(  u_{xx}+u\right)  ^{-2},\quad
P=\mathrm{constant}, \label{tri}%
\end{equation}
with any matrix $P$ of any dimension $n\times n$. However, these matrices $X$
and $T$ (\ref{tri}) commute, $\left[  X,T\right]  =0$, and the corresponding
ZCR (\ref{cxt}) is nothing but $n^{2}$ copies of the evident conservation law
of the equation (\ref{cqe}). In this sense, all the ZCRs sought, with
$X=X\left(  x,t,u\right)  $, turn out to be trivial, up to gauge
transformations (\ref{gxt}) with arbitrary $G\left(  x,t\right)  $.

\section{Recursion operator\label{s4}}

Let us derive a recursion operator of the equation (\ref{cqe}) from the matrix
$X$ (\ref{fmx}) of its ZCR. We do it, following the way described in
\cite{Sak3} (see also references therein). The recursion operator comes from
the problem of finding the class of evolution equations%
\begin{equation}
u_{t}=f\left(  x,t,u,u_{x},\ldots,u_{x\ldots x}\right)  \label{evo}%
\end{equation}
that admit ZCRs (\ref{cxt}) with the predetermined matrix $X$ (\ref{fmx}) and
any $2\times2$ matrices $T\left(  \alpha,x,t,u,u_{x},\ldots,u_{x\ldots
x}\right)  $ of any order in $u_{x\ldots x}$.

The characteristic form of the ZCR (\ref{cxt}) of an equation (\ref{evo}),
with $X$ given by (\ref{fmx}), is%
\begin{equation}
fC=\nabla_{x}S, \label{chf}%
\end{equation}
where $C$ is the characteristic matrix,%
\begin{equation}
C=\frac{\partial X}{\partial u}-\nabla_{x}\left(  \frac{\partial X}{\partial
u_{x}}\right)  , \label{chm}%
\end{equation}
the operator $\nabla_{x}$ is defined as $\nabla_{x}H=D_{x}H-\left[
X,H\right]  $ for any (here, $2\times2$) matrix $H$, and the matrix $S$ is
determined by%
\begin{equation}
S=T-f\frac{\partial X}{\partial u_{x}}. \label{stx}%
\end{equation}

The explicit form of $C$ (\ref{chm}) for $X$ (\ref{fmx}) is%
\begin{equation}
C=%
\begin{pmatrix}
0 & -2\mathrm{i}\alpha^{2}\mathrm{e}^{2\alpha u_{x}}\left(  u_{xx}+u\right)
\smallskip\\
-2\mathrm{i}\alpha^{2}\mathrm{e}^{-2\alpha u_{x}}\left(  u_{xx}+u\right)  & 0
\end{pmatrix}
. \label{efc}%
\end{equation}
Under the gauge transformations (\ref{gxt}) with any $G\left(  \alpha
,x,t,u,u_{x},\ldots,u_{x\ldots x}\right)  $, the characteristic matrix $C$ is
transformed as $C\mapsto GCG^{-1}$ \cite{Mar}, and therefore $\det C$ is a
gauge invariant. We have $\det C=4\alpha^{4}\left(  u_{xx}+u\right)  ^{2}$ in
the case of (\ref{efc}), and this proves that the parameter $\alpha$ cannot be
`gauged out' from $X$ (\ref{fmx}), as well as that the matrix $X$ (\ref{fmx})
cannot be transformed by (\ref{gxt}) into some $X$ containing no derivatives
of $u$.

Computing $\nabla_{x}C$, $\nabla_{x}^{2}C$ and $\nabla_{x}^{3}C$, we find the
cyclic basis to be three-dimensional, $\left\{  C,\nabla_{x}C,\nabla_{x}%
^{2}C\right\}  $, with the closure equation%
\begin{equation}
\nabla_{x}^{3}C=c_{0}C+c_{1}\nabla_{x}C+c_{2}\nabla_{x}^{2}C, \label{clo}%
\end{equation}
where%
\begin{align}
c_{0}  &  =\frac{u_{xxxxx}+2u_{xxx}+u_{x}}{u_{xx}+u}\nonumber\\
&  \quad-9\frac{\left(  u_{xxx}+u_{x}\right)  \left(  u_{xxxx}+u_{xx}\right)
}{\left(  u_{xx}+u\right)  ^{2}}+12\frac{\left(  u_{xxx}+u_{x}\right)  ^{3}%
}{\left(  u_{xx}+u\right)  ^{3}},\label{cc0}\\
c_{1}  &  =4\frac{u_{xxxx}+u_{xx}}{u_{xx}+u}-12\frac{\left(  u_{xxx}%
+u_{x}\right)  ^{2}}{\left(  u_{xx}+u\right)  ^{2}}+4\alpha^{2}\left(
u_{xx}+u\right)  ^{2}-1,\label{cc1}\\
c_{2}  &  =5\frac{u_{xxx}+u_{x}}{u_{xx}+u}. \label{cc2}%
\end{align}

Setting $T$ to be traceless (without loss of generality), we decompose the
matrix $S$ (\ref{stx}) over the cyclic basis as%
\begin{equation}
S=s_{0}C+s_{1}\nabla_{x}C+s_{2}\nabla_{x}^{2}C, \label{cds}%
\end{equation}
where $s_{0}$, $s_{1}$ and $s_{2}$ are unknown scalar functions of
$x,t,u,u_{x},\ldots,u_{x\ldots x}$ and $\alpha$. Substitution of (\ref{cds})
into (\ref{chf}) leads us through (\ref{clo}) to%
\begin{equation}
f=D_{x}s_{0}+c_{0}s_{2},\quad s_{0}=-D_{x}s_{1}-c_{1}s_{2},\quad s_{1}%
=-D_{x}s_{2}-c_{2}s_{2}, \label{fcs}%
\end{equation}
where the function $s_{2}$ remains arbitrary. Then, from (\ref{fcs}) and
(\ref{cc0})--(\ref{cc2}), we obtain%
\begin{equation}
f=\left(  M-\lambda N\right)  r, \label{fmn}%
\end{equation}
where $\lambda=4\alpha^{2}$, $r\left(  \lambda,x,t,u,u_{x},\ldots,u_{x\ldots
x}\right)  =s_{2}$ is any function, of any order in $u_{x\ldots x}$, and the
linear differential operators $M$ and $N$ are%
\begin{align}
M  &  =D_{x}^{3}+5\frac{u_{xxx}+u_{x}}{u_{xx}+u}D_{x}^{2}\nonumber\\
&  \quad+\left(  6\frac{u_{xxxx}+u_{xx}}{u_{xx}+u}+2\frac{\left(
u_{xxx}+u_{x}\right)  ^{2}}{\left(  u_{xx}+u\right)  ^{2}}+1\right)
D_{x}\nonumber\\
&  \quad+\left(  \frac{2u_{xxxxx}+3u_{xxx}+u_{x}}{u_{xx}+u}\right. \nonumber\\
&  \qquad\left.  +4\frac{\left(  u_{xxx}+u_{x}\right)  \left(  u_{xxxx}%
+u_{xx}\right)  }{\left(  u_{xx}+u\right)  ^{2}}-2\frac{\left(  u_{xxx}%
+u_{x}\right)  ^{3}}{\left(  u_{xx}+u\right)  ^{3}}\right)  ,\label{dom}\\
N  &  =\left(  u_{xx}+u\right)  ^{2}D_{x}+2\left(  u_{xx}+u\right)  \left(
u_{xxx}+u_{x}\right)  . \label{don}%
\end{align}

Now, using the expansion%
\begin{equation}
r=r_{0}+\lambda r_{1}+\lambda^{2}r_{2}+\lambda^{3}r_{3}+\cdots, \label{rex}%
\end{equation}
we obtain from (\ref{fmn}) the expression for the right-hand side $f$ of the
represented equation (\ref{evo}), such that $\partial f/\partial\lambda=0$
holds,%
\begin{equation}
f=Mr_{0}, \label{fmr}%
\end{equation}
as well as the recursion relations for the coefficients $r_{i}\left(
x,t,u,u_{x},\ldots,u_{x\ldots x}\right)  $ of the expansion (\ref{rex}),%
\begin{equation}
Mr_{i+1}=Nr_{i},\quad i=0,1,2,\ldots. \label{rrs}%
\end{equation}
The problem has been solved: for any set of functions $r_{0},r_{1}%
,r_{2},\ldots$ satisfying the recursion relations (\ref{rrs}), the expression
(\ref{fmr}) determines an evolution equation (\ref{evo}) admitting a ZCR
(\ref{cxt}) with the matrix $X$ given by (\ref{fmx}).

It only remains to notice that, if a set of functions $r_{0},r_{1}%
,r_{2},\ldots$ satisfies the recursion relations (\ref{rrs}), then the set of
functions $r_{0}^{\prime},r_{1}^{\prime},r_{2}^{\prime},\ldots$ determined by
$r_{i}^{\prime}=N^{-1}Mr_{i}$ ($i=0,1,2,\ldots$) satisfies (\ref{rrs}) as
well. Therefore the evolution equation $u_{t}=f^{\prime}$ with $f^{\prime
}=Mr_{0}^{\prime}=MN^{-1}Mr_{0}=MN^{-1}f=Rf$ admits a ZCR (\ref{cxt}) with $X$
(\ref{fmx}) whenever an equation $u_{t}=f$ does. Eventually, (\ref{dom}) and
(\ref{don}) lead us to the following explicit expression for the recursion
operator $R=MN^{-1}$ of the equation (\ref{cqe}):%
\begin{equation}
R=\frac{1}{u_{xx}+u}D_{x}\frac{1}{u_{xx}+u}\left(  D_{x}+D_{x}^{-1}\right)  .
\label{ero}%
\end{equation}

\section{Conclusion\label{s5}}

Some remarks on the obtained results follow.

We succeeded in transforming the new Chou--Qu equation (\ref{cqe}) into an
integrable equation, the old and well-studied mKdV. The applicability of
Miura-type transformations, however, is not restricted to integrable equations
only. For instance, the original Miura transformation relates very wide
(continual) classes of (mainly non-integrable) evolution equations \cite{Sak4}.

We found the simplest nontrivial ZCR of the evolution equation (\ref{cqe}).
Its matrix $X$ (\ref{fmx}) contains $u_{x}$. For this reason, such a ZCR
cannot be detected by those existent methods, which assume, as a starting
point, that $X=X\left(  x,t,u\right)  $ must suffice in the case of evolution equations.

Of course, we could derive the obtained recursion operator (\ref{ero}) from
the well-known recursion operator of the mKdV through the transformations
found. However, we used a different method instead, mainly in order to
illustrate, by this rather complicated example of $X=X\left(  \alpha
,u,u_{x}\right)  $ (\ref{fmx}), how the method works algorithmically.


\begin{thebibliography}{9}

\bibitem {CQ}K.-S.~Chou and C.~Qu. Integrable equations arising from motions
of plane curves. \textit{Physica D}, 162:9--33, 2002.

\bibitem {Sak1}S.~Yu.~Sakovich. Fujimoto--Watanabe equations and differential
substitutions. \textit{J. Phys. A: Math. Gen.}, 24:L519--L521, 1991.

\bibitem {Sak2}S.~Yu.~Sakovich. On Miura transformations of evolution
equations. \textit{J. Phys. A: Math. Gen.}, 26:L369--L373, 1993.

\bibitem {Wad1}M.~Wadati. The exact solution of the modified
Korteweg--de~Vries equation. \textit{J. Phys. Soc. Jpn.}, 32:1681, 1972.

\bibitem {Wad2}M.~Wadati. The modified Korteweg--de~Vries equation. \textit{J.
Phys. Soc. Jpn.}, 34:1289--1296, 1973.

\bibitem {Sak3}S.~Yu.~Sakovich. Cyclic bases of zero-curvature
representations: five illustrations to one concept. \textit{E-print arXiv},
nlin.SI/0212019, 2002.

\bibitem {Mar}M.~Marvan. A direct procedure to compute zero-curvature
representations. The case $\mathrm{sl}_{2}$. In \textit{Proceedings of the
International Conference on Secondary Calculus and Cohomological Physics},
Moscow, Russia, August 24--31, 1997. \textit{ELibEMS}, http://www.emis.de/proceedings/SCCP97.

\bibitem {Sak4}S.~Yu.~Sakovich. The Miura transformation and Lie--B\"{a}cklund
algebras of exactly solvable equations. \textit{Phys. Lett. A}, 132:9--12, 1988.
\end{thebibliography}
\end{document}